\documentclass[conference]{IEEEtran}
\makeatletter
\def\ps@headings{%
\def\@oddhead{\mbox{}\scriptsize\rightmark \hfil \thepage}%
\def\@evenhead{\scriptsize\thepage \hfil \leftmark\mbox{}}%
\def\@oddfoot{}%
\def\@evenfoot{}}
\makeatother
\IEEEoverridecommandlockouts
\usepackage{cite}
\usepackage{amsmath,amssymb,amsfonts}
\usepackage{algorithmic}
\usepackage{graphicx}
\usepackage{hyperref}
\usepackage{textcomp}
\usepackage{xcolor}
\def\BibTeX{{\rm B\kern-.05em{\sc i\kern-.025em b}\kern-.08em
    T\kern-.1667em\lower.7ex\hbox{E}\kern-.125emX}}

\title{Attack Graph Generation on HPC Clusters
\footnotesize 
\thanks{A preliminary version of this paper appeared in CSCE 2024. 
The final published version is available at Springer, Cham \href{https://doi.org/10.1007/978-3-031-85638-9_9}{https://doi.org/10.1007/978-3-031-85638-9\_9}. 
This is the author’s version and may differ from the published version.}
}

\author{\IEEEauthorblockN{Ming Li}
\IEEEauthorblockA{\textit{Tandy School of Computer Science} \\
\textit{The University of Tulsa}\\
Tulsa, United States \\
ming-li@utulsa.edu}\\\and
\IEEEauthorblockN{John Hale}
\IEEEauthorblockA{\textit{Tandy School of Computer Science} \\
\textit{The University of Tulsa}\\
Tulsa, United States \\
john-hale@utulsa.edu}

}

\date{}

\begin{document}

\maketitle

\begin{abstract}
Attack graphs (AGs) are graphical tools to analyze the security of computer networks.
By connecting the exploitation of individual vulnerabilities, AGs expose possible multi-step attacks against target networks, allowing system administrators to take preventive measures to enhance their network's security.
As powerful analytical tools, however, AGs are both time- and memory-consuming to be generated.
As the numbers of network assets, interconnections between devices, as well as vulnerabilities increase, the size and volume of the resulting AGs grow at a much higher rate, leading to the well-known state-space explosion.
In this paper, we propose the use of high performance computing (HPC) clusters to implement AG generators.
We evaluate the performance through experiments and provide insights into how cluster environments can help resolve the issues of slow speed and high memory demands in AG generation in a balanced way.

\end{abstract}


\maketitle

\section{Introduction}
Attack graphs (AGs) visualize possible paths attackers can take to compromise computer networks \cite{c4}, cyber-physical systems (CPSs) \cite{c20}, IoT \cite{c21}, and even networks of Docker containers \cite{c22}.
AGs allow users to logically connect individual vulnerabilities together to reveal multi-step attacks, which might be unseen if each vulnerability is handled separately.
AGs are generated with input information modeling network assets, interconnections between entities, and vulnerabilities.
The output of AG generators typically consists of node and edge sets, and other relevant information.
The structure of the generated AGs can be analyzed, which identifies nodes, edges and vulnerabilities that are pivotal to achieve attackers' goals.
By further applying probability based approaches \cite{c23}, the likelihood of different attack paths can be compared.
Accordingly, system administrators are informed of more valuable intelligence of the weakness in their system.
They can concentrate the limited time, money and man-power on addressing the most pressing security needs.

The generation of AGs is the most challenging aspect in their application.
Starting from some initial states, the input set of vulnerabilities are repeatedly applied to derive new states.
Most AGs have a tree-like structure.
The farther away from the tree root, the more nodes are branched out.
The earliest AG models, such as \cite{c25} and \cite{c24}, permute all the vulnerabilities to enumerate every possible attack path.
Each AG node in these models represents a network state, which describes the security status of all the network entities.
Each edge is the exploitation of one or more vulnerabilities, and causes a transition between two states.
While these models provide the most detailed security evaluation, they suffer from the exponential growth of the state space as the input size increases \cite{c24}.
To address the issue of state-space explosion, later research proposed more scalable AG models, such as logical AGs \cite{c4,c11}.
In these models, AG nodes are no longer defined to describe the entire network, instead, they may just represent a specific pre- or post-condition, a vulnerability, or a privilege of an attacker on a certain host.
The edges are simply causal connections between nodes and are not associated with any exploitation operations.
Logical AGs and their variations \cite{c26, c27} often assume that attackers will never relinquish a privilege already acquired from previous attack steps, therefore, further reduce the state space to be explored in the generation process.
The generators for Logical AGs and its variations are demonstrated to be polynomial over their input size, which are more efficient than those for state-enumeration AGs.
While the generation complexity of novel AG models are reduced because of simplified model definition and the monotonicity assumptions, they are not completely free of the scalability issue.
When such models are applied to analyze the security posture of large-scale networks, the total computation task and the required memory capacity still easily overwhelm single PCs and small-scale servers.
\cite{c17, c18, c13} introduced parallelism into the AG generation process, however, their efforts are limited in the environment of single computers.
Although distributed AG generation is not a novel idea, to the best of our knowledge, there is no AG generator aiming to run on high performance computing (HPC) clusters, let alone any useful performance data on such.
We observe that AG generation should be treated as other computation intensive tasks and seek the help of HPC.

In this research, we design a parallel algorithm for AG generation that utilizes OpenMPI processes and OpenMP threads to break down the generation task and explore partial state space in parallel.
We conduct the performance evaluation on OSCER, an HPC cluster from University of Oklahoma \cite{c28}.
Our research fills the aformentioned gap and provides design and engineering knowledge to industry and academia that need effective solutions to AG generation.
\section{Related Research}
Research efforts to address the scalability issue of AG models can be partitioned into two tracks.
One track simplifies AG definition to reduce the complexity of the generation process, which is represented logical AGs. 
The other track applies multi-threaded programming to accelerate the exploration of AG state space.
Typical platforms are either single PCs or small servers.
\subsection{Logical AGs}
Two consecutive papers \cite{c4} and \cite{c11} established the foundation of logical AG models.
The nodes in these AGs are categorized as SINK, AND and OR nodes, representing input facts, vulnerability exploitations and derived facts.
As exploitations are defined as a special type of nodes, edges in logical AGs only represent dependence between nodes.
Backtracking is one of the culprits that cause state-space explosion in state-enumeration AG models.
To address this, logical AGs assume that attackers will never relinquish any privileges they have already acquired.
This monotonicity assumption helps eliminate unnecessary permutations of exploitations during state-space exploration, giving logical AGs and its variations \cite{c26, c27} an advantageous polynomial time generation characteristic. 

\subsection{Multi-threaded AG Generation}
Multi-threaded programs are implemented in \cite{c13,c14,c9} to accelerate the generation of AGs.
The data structure to store the resulting AG is shared among the participating threads, which are either OpenMP threads in \cite{c13} or CUDA warps in \cite{c14}.
Each thread starts with a few nodes assigned to it from the initial frontier prepared by a master thread and explores its partial state space.
In \cite{c14}, to take advantage of GPU's computational power, the SIMD threads in each warp further accelerates loops inner to the outer-loop that expand AG nodes.
In \cite{c9}, work-stealing is proposed to balance the workload among the threads, which further reduces the execution time.
While these designs are able to accelerate AG generation, they are implemented on either a single PC or a small server, and the performance worsens sharply as the memory demands exceed the available capacity.

Our research extends the multi-threaded scheme by proposing an AG generation algorithm targeting HPC clusters.
Modern HPC clusters have ample memory on each node, satisfying the needs of many memory-intensive programs.
To the best of our knowledge, however, no existing research ever deployed AG generators on HPC clusters.

In \cite{c17}, a distributed AG generator is proposed.
The authors apply reachability hyper-graph partitioning to divide the target network into groups of networked software applications.
Each group is assigned as a task to a search agent to derive a part of the AG.
The multiple agents communicate with one another through TCP sockets and access a virtually shared memory to avoid redundant expansion of nodes that are already processed.
While this AG generator is designed to execute with distributed computing agents, the experiments yielding a speedup of X2.65 were actually conducted on a single computer with a quad-core Intel processor.
More experiments are needed to evaluate if this distributed AG generator can perform equally well on a real distributed platform, especially after adding the overhead from TCP/IP communication between search agents and from accessing the virtually shared memory. 

In \cite{c18}, the authors proposes to use cluster-computing environment Spark to parallelize AG generation. 
Utilizing multiple Spark executors enables each to generate a distinct sub-AG.  
Following parallel execution, these subgraphs are merged into a comprehensive resulting AG.
To optimize the parallelization efficiency, a multilevel k-way partition algorithm divides the input network into smaller segments according to topology, which significantly reduces the workload added to each executor.
While the experiments in \cite{c18} on a single computer indicate that the Spark-based scheme outperforms a distributed AG generation algorithm, crucial experiment details, such as the implementation specifics of the baseline AG generator being compared, and the dimensions of each target AG, are omitted.

Different from these existing efforts to parallelize AG generation, we not only deploy our parallel AG generator crossing multiple nodes on an HPC cluster, but also tune the platform parameters to examine the impacts of the hardware configuration on the performance and cost.
Furthermore, we profile the execution times of different components in our AG generator to identify the most critical one and propose further optimizations to speed it up.

\section{AG Generation On HPC Clusters}
This section introduces the AG model used by this paper and applies it to an example network. 
It then presents a parallel algorithm for the model's AG generator to be deployed on HPC clusters. 
\subsection{AG Model}
Our AG model follows the design in \cite{c9}.
The model defines an AG as a tuple:
\begin{equation}
AG = \{V, E\}
\end{equation}, where V is the set of nodes and E is the set of directed edges.
Each node represents the set of properties of network assets relevant to attacks.
Each edge represents the exploitation of one or more vulnerabilities, causing a state transition from one node to the other. 
To build an AG for a target network, the input must include:
\begin{itemize}
\item A list of assets, which encompasses network devices and software entities. 
\item A list of vulnerabilities. 
Each vulnerability is formatted as a set of pre-conditions and a set of post-conditions.
\item A set of initial properties of network assets, which essentially defines the root node of the AG.
According to \cite{c9}, the AG tree structure originates from the root node, and all derived nodes are either intermediate states or target states after attacks are carried out successfully.
\end{itemize}
\begin{figure}[t]
\centerline{\includegraphics[width=0.40\textwidth]{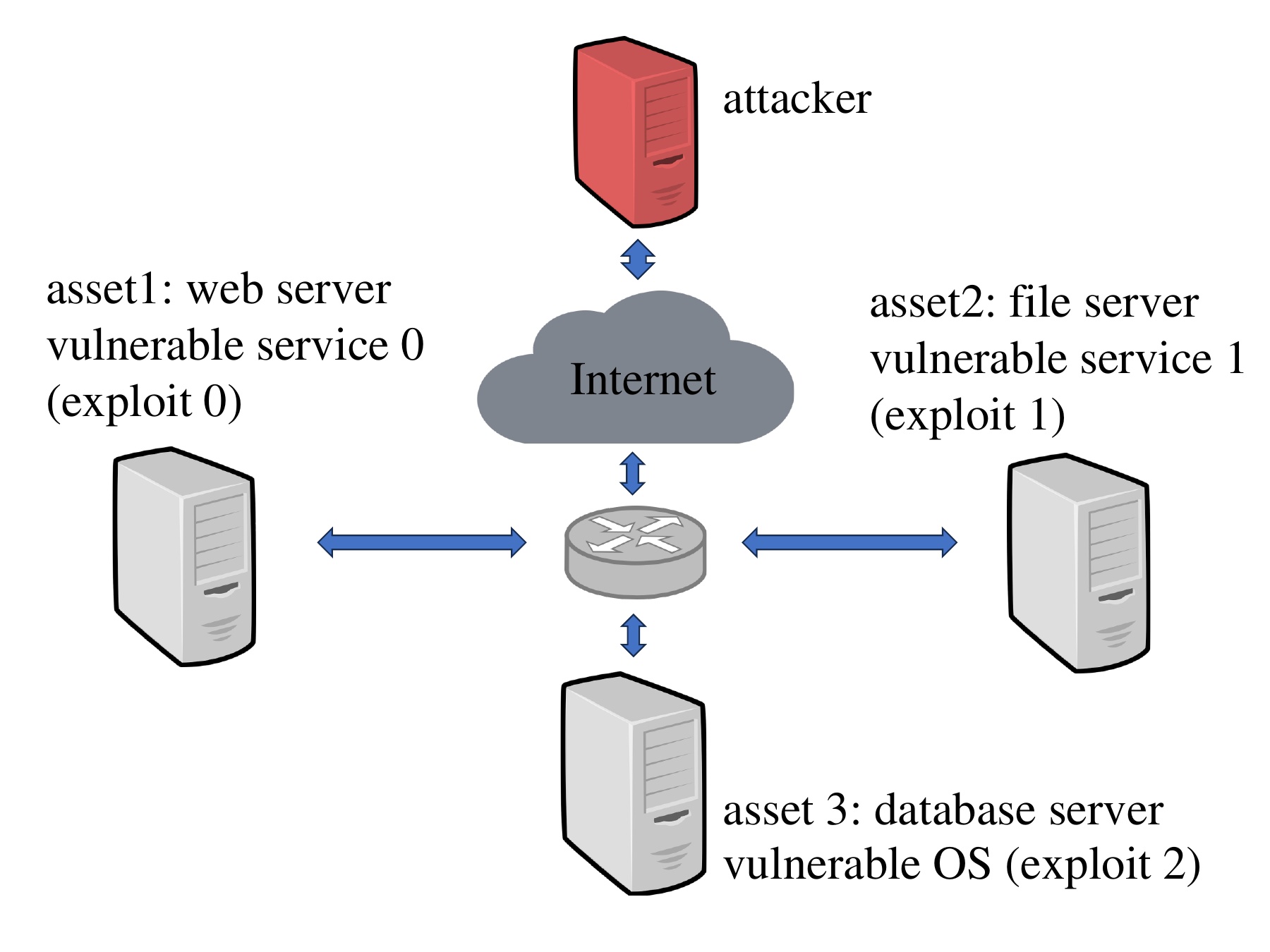}}
\caption{A target network for AG generation.}
\label{pic1} 
\end{figure}

As an example, Fig.\ref{pic1} shows a small network with three servers.
The security policy regulates that any user from the Internet can only use the web or file service.
The database server only provides backend service to the other two servers.
The web server and the file server both have vulnerable services that might be used by an attacker to gain root privileges.
In addition, the database server has a bug in its OS, which might be exploited by an attacker that has a foothold on either the web server or the file server. 
\begin{figure}[t]
\centerline{\includegraphics[width=0.4\textwidth]{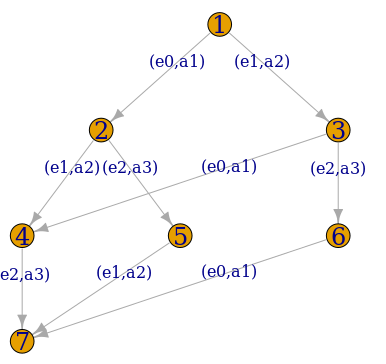}}
\caption{Generated AG for the network in Fig.\ref{pic1}.}
\label{pic2} 
\end{figure}
With the given input information, the AG is generated as in Fig.\ref{pic2}.
Considering that the attacker may choose the database server as the final goal, the AG shows that the exploitation 2 on the database server always conditions on either exploitation 0 on the web server or exploitation 1 on the file server.
Thus, this AG helps security administrator to identify all the possible multi-step attacks that can compromise the database server.

\subsection{Parallel AG Generator on HPC Clusters}
\begin{figure}[tb]
\centerline{\includegraphics[width=0.52\textwidth]{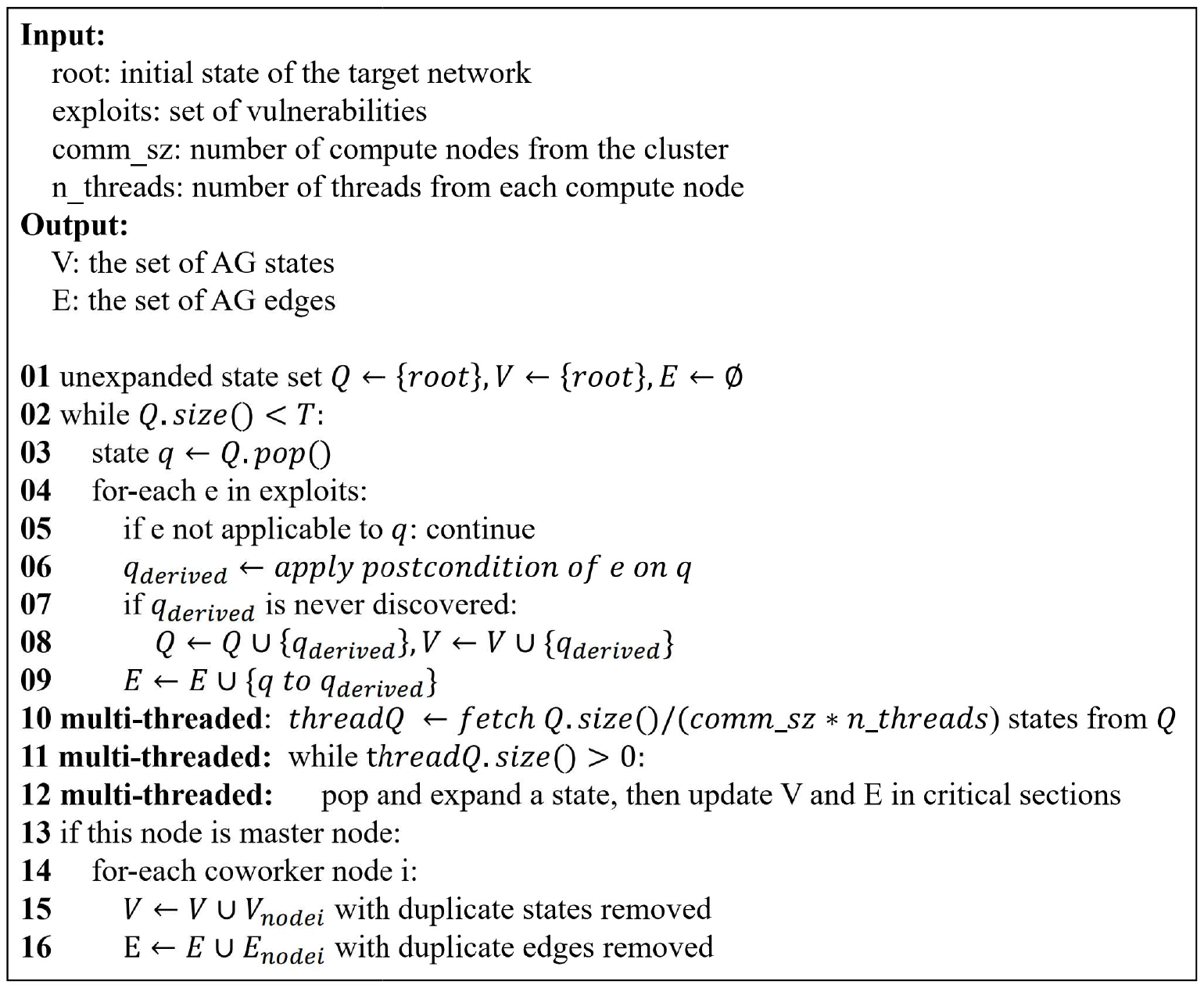}}
\caption{Parallel algorithm to generate AG on HPC clusters.}
\label{pic3} 
\end{figure}
To generate AGs on HPC clusters, we design a parallel algorithm as illustrated in Fig.\ref{pic3}, which comprises three distinct phases.
In phase 1 (lines 1-9), the initial AG state (Note: AG state is used hereafter instead of AG node to avoid confusing with cluster node) and its derived states are expanded by each cluster node locally to fill a per-node queue (Q) with more unexpanded states.
The queue is identical on each cluster node as they take identical input.
When the size of the per-node queue grows greater than a preset threshold T, the multi-threaded phase 2 (lines 10-12) begins.
Suppose comm\_sz is the number of cluster nodes and n\_threads is the number of threads from each node, then the total number of threads in the multi-threaded phase is comm\_sz*n\_threads.
Each thread maintains its own thread queue (threadQ), with an initial size equal to the per-node queue size (Q.size()) divided by the total number of threads.
The partition of the unexpanded states in the per-node queue is cyclic, aiming to divide the AG state space evenly among all the threads.
For instance, with 9 states (s1-s9 in discovered order) in the per-node queue to begin with, a parallel AG generator launched on three cluster nodes (n1-n3) and each with three threads (t1-t3) will partition the initial work as follows:
\begin{verbatim}
n1-t1: s1, n2-t1: s2, n3-t1: s3,
n1-t2: s4, n2-t2: s5, n3-t2: s6,
n1-t3: s7, n2-t3: s8, n3-t3: s9
\end{verbatim}
As inter-node communication is more expensive than local computations, in phase 1 and 2, each cluster node explores its partial state space independently.
In phase 3 (lines 13-16), a master node merges all the partial graphs into a complete AG.
The merging needs to remove duplicate nodes and edges through hashing methods.
With comm\_sz nodes, the merging requires a total of comm\_sz-1 inter-node communications, which might add a long latency to the total execution time.
If no merging is required, however, each node can keep its partial AG in the local memory or store it into an AG database.

\section{Performance Evaluation}
We implement the parallel algorithm for AG generation on HPC clusters with the hybrid of Message Passing Interface (MPI) and Open Multi-Processing (OpenMP), both of which are available in most HPC environments.
Specifically, each cluster node in the algorithm is embodied by an OpenMPI process.
Each OpenMPI process forks multiple OpenMP threads for the multi-threaded phase in the algorithm.

\subsection{Performance Evaluation}
\begin{figure}[tb]
\centerline{\includegraphics[width=0.48\textwidth]{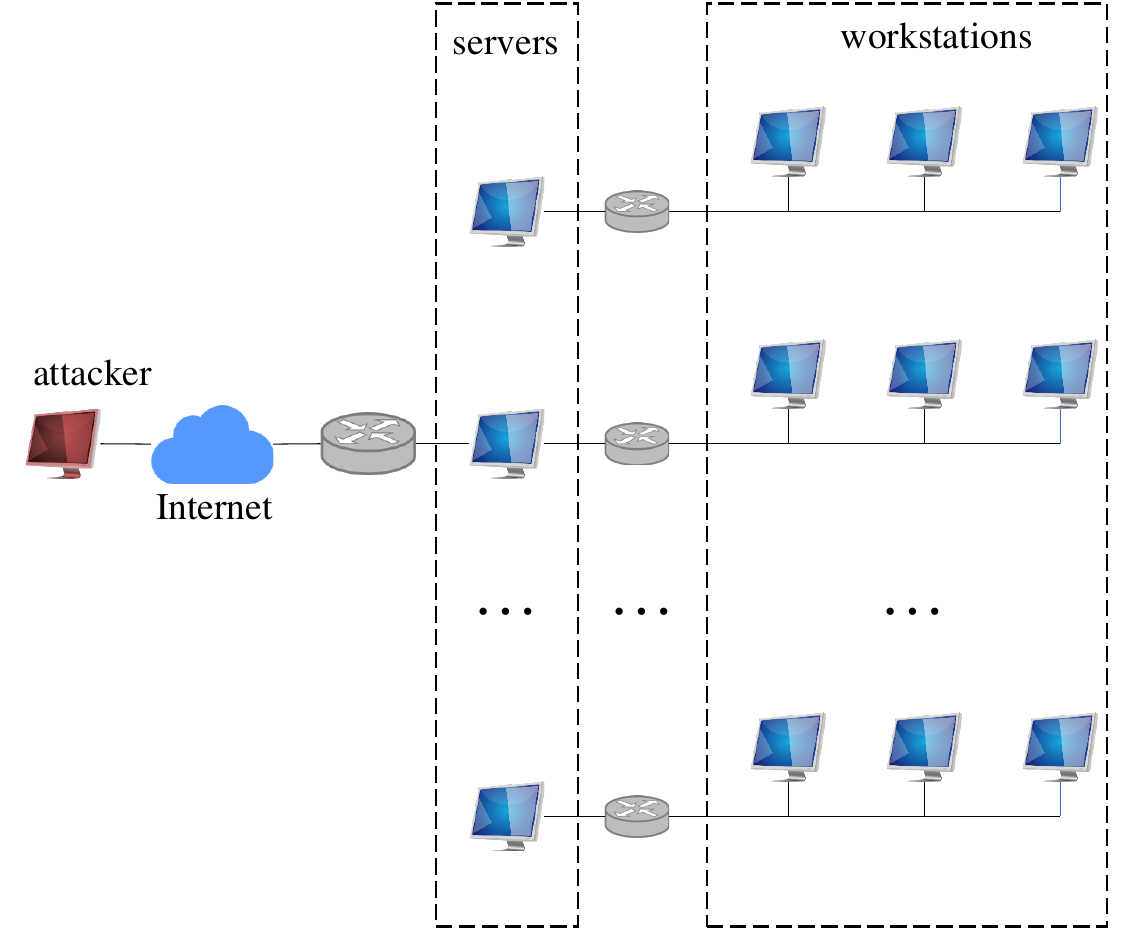}}
\caption{Target network for AG generation in performance evaluation.}
\label{pic4} 
\end{figure}
The AG to be generated targets a network structured in a tree topology in Fig.\ref{pic4}.
The attacker from the Internet has the option to compromise any of the servers.
With a compromised server as a foothold, the attacker can subsequently attack any of the workstations connected to the server via LAN.
The target network has 150 computers, 20\% of which have a vulnerability to be exploited. 
The generated AG has 5,859,375 states and 56,640,625 edges.
The storage cost is 13.5GB, which is not small compared with the limited memory capacity of single computers.

As the baseline configuration, two compute nodes are tested first, each with the number of threads per-node tuned from 2 to 40, which matches the maximum number of CPUs per-node.
\begin{figure}[t]
\centerline{\includegraphics[width=0.50\textwidth]{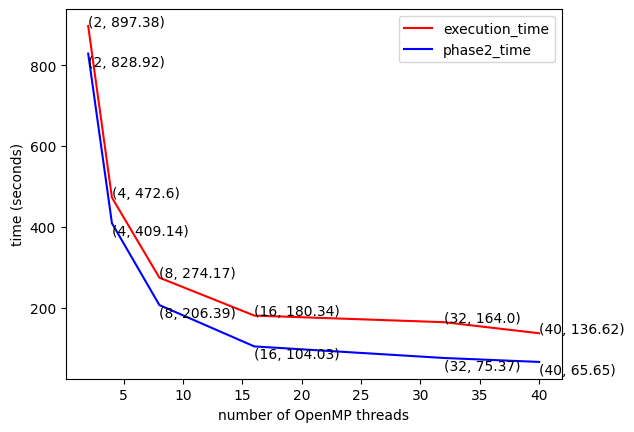}}
\caption{Baseline performance with two compute nodes. Each node has 40 CPUs (Intel Xeon E5-2650 @2.3Ghz) and one 32GB memory.}
\label{pic5} 
\end{figure}
Fig.\ref{pic5} shows that as the number of threads in the multi-threaded phase increases, the total execution time decreases.
Considering only phase 2, its execution time reduces about 50\% each time the number of threads doubles, demonstrating the effectiveness of intra-node parallelism in accelerating the AG generation process. 
Based on the algorithm in Fig.\ref{pic3}, there is only one MPI send/recv communication between node 0 (master) and node 1 during phase 3 to merge partial AGs.
For the case of 40 OpenMP threads per-node, the breakdown of the execution time:
\begin{verbatim}
phase 1 total: 0.48 seconds
phase 2 total: 65.65 seconds
Phase 3 total: 69.47 seconds
- MPI comm prep: 0.65 seconds
- MPI send/recv: 23.95 seconds
- Merging states: 2.58 seconds
- Merging edges: 42.29 seconds
total time: 136.62 seconds
\end{verbatim}
Phase 1 takes a very short duration to prepare the workload for the multi-threaded phase 2.
However, phase 3 contributes the largest amount to the total execution time.
While the time for MPI send/recv is a necessary cost (23.95 seconds) for the AG merging process, merging edges takes a longer time of 42.29 seconds, implying that further optimization is required.

The next experiment tunes the number of invoked compute nodes.
Specially, the execution times of the three phases are compared under 2-node, 3-node and 4-node settings.
\begin{figure}[t]
\centerline{\includegraphics[width=0.45\textwidth]{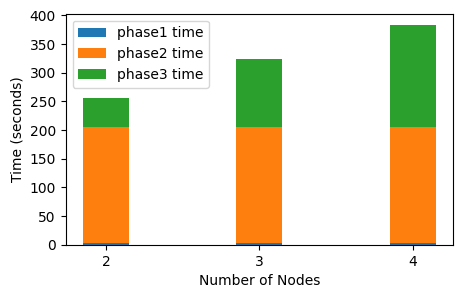}}
\caption{Tune the number of compute nodes. Node configuration: 8 OpenMP threads; 32 GB memory per-node; CPU-Intel Xeon E5-2650 @2.3Ghz.}
\label{pic6} 
\end{figure}
As Fig.\ref{pic6} suggests, invoking more compute nodes increases the overall execution time.
From 2 nodes to 4 nodes, the three settings spend approximately equal time on phase 1 and 2.
However, phase 3 becomes more expensive.
This trend is attributed to a larger overhead on MPI send/recv communication between compute nodes.
For example, under 4-node settings, node 1, 2 and 3 must send their AG states and edges (essentially the partial AG itself) to node 0 for merging.
In an application setting where merging partial AGs is not required, the time of phase 3 will not be a factor.
As a result, the total execution time is expected to be non-increasing even if more compute nodes are enlisted. 

On the other hand, although the total running time is unsatisfactory due to the implementation of phase 3, the benefits of running with multiple HPC nodes to mitigate the pressure of memory usage should not be overlooked.
Our experimental AG needs 13.5GB of storage.
A typical OSCER compute node provides about 30GB memory, which can be shared by at most 2 of our MPI processes.
Mapping more processes to each node results in unsuccessful launching of the MPI program. 
With these processes more sparsely distributed, such as one process per-node in the conducted experiments, a master node easily allocates space for the storage of two partial AGs: one for itself and the other for buffering the AG received from other nodes.
Execution time is not the only goal of an efficient solution to AG generation.
Designing an AG generator to run on HPC clusters will need to strike a balance between minimizing execution time and distributing storage cost on participating nodes.

\subsection{Optimization}
To reduce overhead of merging AGs, we propose the following options to accelerate phase 3 of our AG generator.
\begin{itemize}
\item Option 1: create multiple threads in the master node, and each thread receives and merges one partial AG.
This option means each thread in the master node must have its own buffer to store a partial AG, which results in a higher memory demand on the master node.
\item Option 2: create a software pipeline on the master node with multiple threads.
Some threads serve as producers only receiving new partial AGs, while others as consumers only merge those already received into the complete AG. 
\item Option 3: create a hierarchical merging process, which aims to accelerate when more than four nodes are invoked. For instance, the even ID-ed processes merge into the adjacent odd ID-ed first, then the odd ID-ed ones merge together to build the complete AG.
\end{itemize}
As of the writing of this paper, experiments are underway to evaluate these optimization options. 
Option 3 is tested with the same cluster setting as the experiments already conducted. Both option 1 and 2 require more memory than the current implementation, and they are being tested on OSCER's large memory queues.  

\section{Conclusions and Future Work}
To address the scalability issue of AG generation, this paper presents a parallel algorithm and implementation on HPC clusters.
The proposed algorithm partitions the generation process into three phases. 
Phase 1 runs a single thread per-node to prepare enough workload for multiple threads. 
Phase 2 runs a hybrid of MPI processes and OpenMP threads to accelerate the exploration of partial AGs. 
Phase 3 merges partial AGs into a complete one through MPI communications.
The experimental results reveal that AG generation on HPC clusters can achieve an equilibrium between accelerating the generation process and reducing the memory demands on the computing device.
For subsequent research, we will complete the design and experiments that optimize the merging of partial AGs.
In addition, we will explore possible solutions to eliminate the need to merge partial AGs on a master node, for instance, via building a distributed database to store AG nodes and edges.

\section*{Acknowledgment}
The computing of this project was performed at the OU Supercomputing Center for Education \& Research (OSCER) at the University of Oklahoma (OU). OSCER Research Computing Facilitator Thang Ha provided valuable technical expertise.

\bibliographystyle{IEEEtran}
\bibliography{Attack_Graph_Generation_on_HPC_Clusters}

\end{document}